\documentclass[aps,pre,twocolumn,superscriptaddress,showpacs]{revtex4}
\usepackage{amsmath,latexsym,epsfig,times} 

\begin{document}

[Phys. Rev. E {\bf 66}, 065102 (2002)]

\title{Cascade-based attacks on complex networks}

\author{Adilson E. Motter}
\email{motter@chaos3.la.asu.edu}
\affiliation{Department of Mathematics, CSSER,
Arizona State University, Tempe, Arizona 85287}

\author{Ying-Cheng Lai}
\affiliation{Department of Mathematics, CSSER,
Arizona State University, Tempe, Arizona 85287}
\affiliation{Departments of Electrical Engineering and Physics,
Arizona State University, Tempe, Arizona 85287}

 
\begin{abstract}

We live in a modern world supported by large, complex networks.
Examples range from financial markets to 
communication and transportation systems.
In many realistic situations the flow
of physical quantities in the network, as characterized by the loads on nodes,  
is important. 
We show that for such networks where loads can redistribute among the nodes, 
intentional attacks can lead to a {\it cascade} of overload failures,
which can in turn cause the entire or a substantial part of the network to collapse.
This is relevant for   
real-world networks that possess a highly heterogeneous distribution of {\it loads}, 
such as the Internet and power grids. 
We demonstrate that
the heterogeneity of these networks makes them particularly vulnerable to attacks
in that a large-scale cascade may be triggered by disabling a {\it single} key node.
This brings obvious concerns on the security of such systems.

\end{abstract}
\pacs{89.20.Hh, 89.75.-k, 05.10.-a}
\maketitle

Complex networks are an essential part of a modern society \cite{Strogatz:2001,AB:2002}. 
It has been shown that many networks, such as the world-wide web (WWW), 
the Internet, and electrical power grids, present a
surprisingly small average distance between nodes and a highly 
organized distribution of links per node \cite{WS:1998,BA:1999,AJB:2000}. 
Generally, the average distance will not be affected by the 
removal of a random subset of nodes, but it will increase
significantly if the removed nodes are among the most connected ones \cite{AJB:2000}
(see also Refs. \cite{cohen:2000,callaway:2000,cohen:2001}).
The existence of a giant connected component in the network, however, 
does not depend on the presence of highly connected nodes. For instance,
the WWW has homepages with many thousands of hyperlinks and can remain well
connected after the removal of all homepages with five or more
hyperlinks \cite{Broder:2000}.
In addition, the giant component itself is typically a 
{\it small-world network} \cite{Watts:book} even after
the removal of all highly connected nodes \cite{Alessandro:2002}. 
These pioneering studies on network security address mainly static 
properties, $i.e.$, the effect of different network architectures.
They suggest that the network connectivity, and hence its functionability, 
is robust against random failure of nodes \cite{AJB:2000,cohen:2000,callaway:2000}
and to some extent is even robust against intentional attacks 
\cite{Broder:2000,Alessandro:2002}.
Here we show that for many physical networks, the removal of nodes
can have a much more devastating consequence when the intrinsic {\it dynamics} 
of flows of physical quantities in the network is taken into
account. In a power transmission grid, for instance, 
each node (power station) deals with a load of
power. The removal of nodes, either by random breakdown or intentional attacks, 
changes the balance of flows and leads to a global redistribution of loads
over all the network. This can trigger a cascade of overload
failures \cite{Watts:2002,Moreno:2002},
as the one that happened on August 10, 1996 in the western United States 
power grid \cite{Carreras:2000,Sachtjen:2000}.
Another example is the Internet \cite{pastor:2001,willinger:2002,goh:2002},
where the load represents the amount of information a node (router)
is requested to transmit per unit of time, and overloads correspond to 
congestion \cite{congestions}.
Internet collapses caused by congestion have been reported 
since its very beginning \cite{Jacobson:1998}.
In this Rapid Communication, we introduce a model for cascading failure in complex networks
and show that it is applicable to realistic networks such as the Internet and power grids.

For a given network, suppose that at each time step
one unit of the relevant quantity, which can be information, energy,
etc., is exchanged between every pair of nodes 
and transmitted along the shortest path connecting them.
The load at a node is then the total number of shortest 
paths passing through the node \cite{Newman:2001,Goh:2001,HK:2002}.
The capacity of a node is the maximum load that the node can handle.
In man-made networks, the capacity is severely limited by cost. 
Thus, it is natural to assume that the capacity
$C_j$ of node $j$ is proportional to its initial load $L_j$,
\begin{equation}
C_j=(1+\alpha)L_j,\;\;\;  j=1,2,...N,
\label{capacity}
\end{equation}
where the constant $\alpha\geq 0$ is the {\it tolerance} 
parameter, and $N$ is the initial number of nodes.
When all the nodes are on, the network operates in a free-flow 
state insofar as $\alpha\geq 0$.
But, the removal of nodes in general changes the 
distribution of shortest paths. The load at a particular
node can then change. If it increases and becomes larger 
than the capacity, the corresponding node
fails. Any failure leads to a new redistribution of loads and, 
as a result, subsequent failures can occur.
This step-by-step process is what we call a {\it cascading failure},
or a cascade. It can stop after a few steps but it can also propagate and 
shutdown a considerable fraction of the whole network \cite{comment}.
A fundamental question is: 
under what conditions can such a global cascade take place?

Here we focus on cascades triggered by the removal of a single node.
If a node has a relatively small load, its removal will not cause major changes
in the balance of loads, and subsequent overload failures are unlikely to occur.
However, when the load at the node is relatively large,
its removal is likely to affect significantly loads at other nodes
and possibly starts a sequence of overload failures. Our result is 
the following: global cascades occur if 
(1) the network exhibits a highly heterogeneous distribution of loads;
(2) the removed node is among those with higher load.
Otherwise, cascades are not expected.
The distribution of loads is in turn highly correlated with the distribution of links:
networks with heterogeneous distribution of links are expected to be heterogeneous
with respect to load so that on average,
nodes with larger number of links will have higher load \cite{Goh:2001}.
This result confirms the robust-yet-fragile property of heterogeneous
networks, which was first observed in Ref. \cite{AJB:2000} for the attack on {\it several} nodes.
The cascade effect is important, however, because a large damage can
be caused in this case by the attack on a {\it single} node.
While a network with more links can be more resistant against cascading failures,
in practice the number of links is limited by cost.

Now we provide evidence for our result. 
We study cascades triggered by random breakdown and by intentional attacks.
To simulate the former, we choose a trigger at random
among all the nodes of the network, as can occur in networks such as power grids
\cite{Carreras:2000}. In the case of attack the
targeted node is selected from those with highest loads
or largest {\it degrees} (number of links at a node).
We consider heterogeneous networks with 
algebraic (scale-free) distribution $P$ of links,
as observed in real systems \cite{Redner:1998,FFF:1999,BA:1999,AB:2002},
\begin{equation}
P(k)\sim k^{-\gamma},
\label{distribution}
\end{equation}
where $k$ denotes the degree and $\gamma$ the scaling exponent,
and compare them with an equivalent homogeneous configuration.
These networks are generated according to the
procedure in Refs. \cite{NSW:2001,Motter:2002}, where the nodes are connected randomly
for a given degree distribution, and self- and repeated links are forbidden.
The damage caused by a cascade is quantified in terms of the
relative size $G$ of the largest connected component,
\begin{equation}
G=N'/N,
\label{size}
\end{equation}
where $N$ and $N'$ are the numbers of nodes
in the largest component before and after the cascade, respectively.

\begin{figure}[t]
\begin{center}
\epsfig{figure=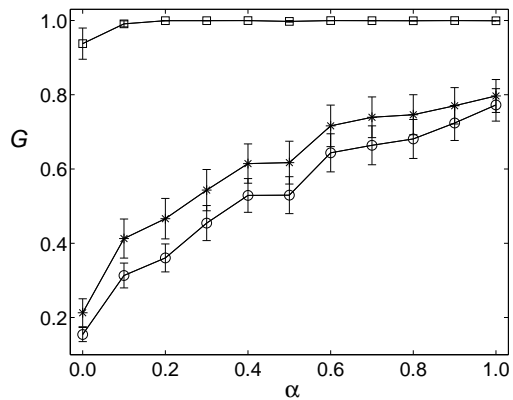,width=6.8cm}
\caption{
Cascading failure in scale-free networks,
as triggered by the removal of 
a single node chosen at random (squares),
or among those with largest degrees (asterisks) or highest loads (circles),
where
$\alpha$ is the tolerance parameter and
$G$ is the relative size of the largest connected component.
Each curve corresponds to the average 
over 5 triggers and 10 realizations of the network.
The error bars represent the standard deviation.
The networks are generated according to the algebraic 
distribution (\ref{distribution}).
For the computations shown we set $\gamma=3$ and $5000\leq N\leq 5100$.
The average degree in the largest component is $\langle k\rangle \approx 2.0$.}
\label{fig1}
\end{center}
\end{figure}

Figure \ref{fig1} shows the relative size $G$ of the largest component
after cascading, as a function of the tolerance parameter $\alpha$, for
a scale-free network. We can see that on average $G$ remains 
close to unity in the case of random breakdowns but it is
significantly reduced under intentional attacks, 
even for $\alpha$ unrealistically large. Indeed, 
the size of the largest component is reduced by more 
than $20\%$ for $\alpha=1$, $i.e.$, for a capacity as large as two times the 
capacity required for the system to operate when all the nodes function normally.
This result is in agreement with intuition, because in the case of random breakdown the
trigger is probably one of the many nodes with small load, while in the case of intentional attack
it is a node with very large load. 
The damage is larger for smaller values of $\alpha$, as it is for load-based attacks when compared with
degree-based attacks. For instance, in the load-based attack for $\alpha=0.2$,
more than $60\%$ of the nodes are affected. For the 5000-node networks used in our simulations, 
it means that a cascade triggered by the attack
on a single node shuts down and disconnects more than 3000 others!

\begin{figure}[b]
\begin{center}
\epsfig{figure=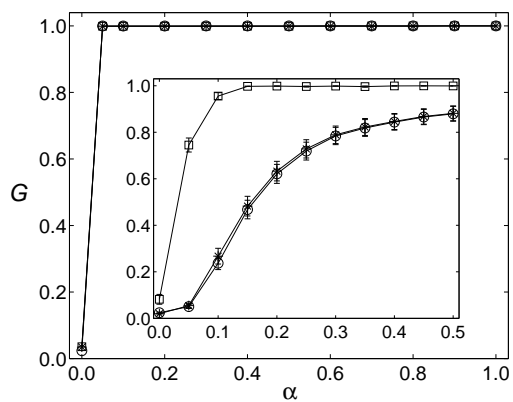,width=6.8cm}
\caption{
Cascading failure in homogeneous networks.
All nodes are set to have the same degree $k= 3$ and $N=5000$.
In the inset, the networks are generated according to the algebraic distribution
(\ref{distribution}) for $k\geq 2$, $\gamma=3$, and $N=5000$.
The resulting average degree is $\langle k\rangle \approx 3.1$.
The legends and other parameters are the same as in Fig. \ref{fig1}.}
\label{fig2}
\end{center}
\end{figure}
 
Figure \ref{fig2} shows the corresponding results for
a homogeneous network with the same number of nodes and  exactly 3 links per node.
To make a meaningful comparison we display in the inset results for
an algebraic network with about the same average degree (actually larger, which strengthens our conclusions).
The homogeneous network does not experience cascading failures due either to
random breakdown or to intentional attacks for $\alpha$ as small as $0.05$.
For the heterogeneous (scale-free) network, for the 
same value of $\alpha$, cascades triggered by the attack on a key node
can reduce the largest connected component to less than 
$10\%$ of the original size, as shown in the inset.
Therefore, homogeneous networks appear to be  
more robust against attacks than the heterogeneous ones.
This conclusion does not rely on the particularities 
of these models, as the same was also observed for
classes of networks with exponential and Poisson-like 
distributions of degrees ($e.g.$, the Erd\"os-R\'enyi model \cite{ER}):
their homogeneity makes them relatively resistant
to cascades triggered by attacks. The networks corresponding 
to the inset of Fig. \ref{fig2} are
generated according to the same scaling distribution of those in Fig. \ref{fig1},
except that in this case the minimal number of links at a node is set to be 2.
Therefore, this inset shows that the fragility of scale-free networks
is due to their heterogeneity and does not rely on the presence of nodes with degree one,
which are easily disconnectable.
Naturally, the increase of the average degree  
reduces the damage of the cascade, as can be seen from a comparison between
Fig. \ref{fig1} and the inset of Fig. \ref{fig2}.

\begin{figure}[b]
\begin{center}
\epsfig{figure=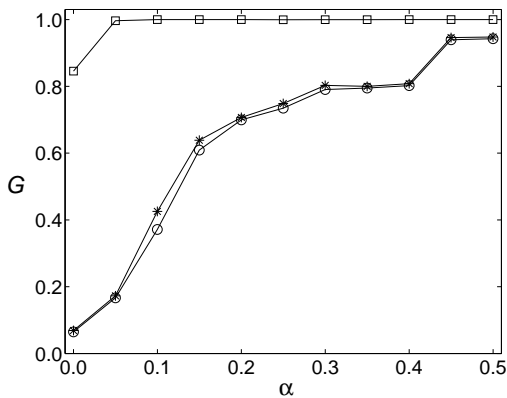,width=6.8cm}
\caption{
Cascading failure in the Internet at autonomous system level \cite{Internet}.
The network has $N=6474$ nodes and $\langle k\rangle \approx 3.88$ links per node, on average.
Each curve corresponds to the average over 5 triggers for attacks and 50 for random breakdown.
The legends are as defined in Fig. \ref{fig1}.}
\label{fig3}
\end{center}
\end{figure}

\begin{figure}[t]
\begin{center}
\epsfig{figure=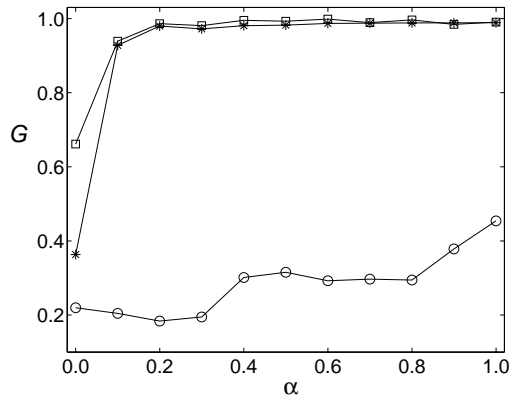,width=6.8cm}
\caption{
Cascading failure in the Western U.S. power transmission grid \cite{power-grid},
which has $N=4941$ and $\langle k\rangle \approx 2.67$.
The average is obtained via 5 triggers for attacks and 50 for random breakdown.
The legends are the same as in Fig. \ref{fig1}.}
\label{fig4}
\end{center}
\end{figure}

Many real-world networks are heterogeneous and 
as such are expected to undergo large-scale cascades
if some vital nodes are attacked, but rarely in the case of random breakdown.
As an example we consider the Internet at autonomous system level \cite{Internet},
which displays an algebraic distribution of links \cite{AJB:2000}.
The damage caused by triggers of higher load or degree is much larger than
that by random breakdown, as shown in Fig. \ref{fig3}.
The cascading failures are rarely triggered by random breakdown
for $\alpha > 0.05$, but more than $20\%$ of the nodes can be disconnected 
with the intentional attack on only one node for $\alpha\leq 0.4$. 
We have also considered the electrical power grid of 
the western United States \cite{power-grid}.
The degree distribution in this network is consistent with an exponential \cite{Amaral:2000}
and is thus relatively homogeneous.
The distribution of loads, however, is more skewed than that displayed
by semirandom networks \cite{NSW:2001,Motter:2002} 
with the same distribution of links, indicating that the
power grid has structures that are not captured by these models.
As a result, global cascades can be triggered by 
load-based intentional attacks but not by random or degree-based removal of nodes,
as shown in Fig.  \ref{fig4}.
We see that the attack on a single node with large load reduces
the largest connected component to less than a half of its initial size,
even when the network is highly tolerant ($e.g.$, $\alpha=1$).

Our result is thus that real networks are naturally 
evolved to be quite resistant to random failure of nodes, 
but the presence of a few nodes with exceptionally {\it large load},
which is known to be ubiquitous in natural and man-made networks,
has a disturbing side effect: the attack on a single important node
(one of those with high load) may trigger a cascade of 
overload failures capable of disabling the network almost entirely.
Such an event has dramatic consequences on the network performance,
because the functionability of a network relies on the ability of the nodes
to communicate efficiently with each other.
What is the use, say, of having a phone if you cannot call anybody?

We conclude with some thoughts on the meaning of our results for security.
An effective attack relies on identifying
vulnerabilities and is far from being random. 
Our society is geographically distributed in
a way that natural hazards are by no means random \cite{editorial}.
An example is the crowding of people, communication, 
transportation, and financial centers
around seismic areas, like the Pacific Rim.
Natural disasters and intentional attacks can then have devastating 
consequences on the complex networks underlying the society.
These consequences will be  more severe if the damage on one 
or few nodes is capable of spreading over the entire network. 
In this sense a cascade-based attack can be much more
destructive than any other strategies of attack previously considered
\cite{AJB:2000,Broder:2000,Motter:2002,callaway:2000,cohen:2001,sole:2001,jeong:2001,HKYH:2002}.

The authors thank R\'eka Albert and Duncan J. Watts for providing
the Internet and power-grid data, respectively.
This work was supported by AFOSR under Grant No. F49620-98-1-0400
and by NSF under Grant No. PHY-9996454.

\end{document}